\documentclass[prd,superscriptaddress,nofootinbib,showpacs]{revtex4}
\usepackage{float}
\usepackage{amssymb,latexsym}
\usepackage{amsmath,amsbsy,bbm}
\usepackage{epsfig,bm}
\usepackage{graphicx,comment}
\usepackage{color}
\long\def\symbolfootnote[#1]#2{\begingroup%
\def\thefootnote{\fnsymbol{footnote}}\footnote[#1]{#2}\endgroup}

\begin{document} 

\title{Subtlety of Studying the Critical Theory of a Second Order Phase Transition}

\author{F.-J. Jiang}
\email[]{fjjiang@ntnu.edu.edu}
\affiliation{Department of Physics, National Taiwan Normal University, 88, Sec.4, Ting-Chou Rd., Taipei 116,
Taiwan}
\author{U. Gerber}
\email[]{gerberu@itp.unibe.ch}
\affiliation{Albert Einstein Center for Fundamental Physics, 
Institute for Theoretical Physics, Bern University, 
Sidlerstrasse 5, CH-3012 Bern, Switzerland}
\vspace{-1cm}
  
\begin{abstract}
We study the quantum phase transition from a super solid phase to a 
solid phase of $\rho = 1/2$ for the extended Bose-Hubbard model on the 
honeycomb lattice using first principles Monte Carlo calculations.
The motivation of our study is to quantitatively understand the impact of 
theoretical input, in particular the dynamical critical exponent $z$, in 
calculating the critical exponent $\nu$. Hence we have carried out four sets 
of simulations with $\beta = 2N^{1/2}$, $\beta = 8N^{1/2}$, $\beta = N/2$, and 
$\beta = N/4$, respectively. Here $\beta$ is the inverse temperature and $N$ is
the numbers of lattice sites used in the simulations. By applying data collapse to 
the observable superfluid density $\rho_{s2}$ in the second spatial direction,
we confirm that the transition is indeed governed by the superfluid-insulator 
universality class. However we find it is subtle to determine the precise 
location of the critical point. For example, while the critical chemical 
potential $(\mu/V)_c$ occurs at $(\mu/V)_c = 2.3239(3)$ for the data obtained 
using $\beta = 2N^{1/2}$, the $(\mu/V)_c$ determined from the data simulated 
with $\beta = N/2$ is found to be $(\mu/V)_c = 2.3186(2)$. Further, while a 
good data collapse for $\rho_{s2}N$ can be obtained with the data determined 
using $\beta = N/4$ in the simulations, a reasonable quality of data collapse 
for the same observable calculated from another set of simulations with 
$\beta = 8N^{1/2}$ can hardly be reached. Surprisingly, assuming $z$ for this 
phase transition is determined to be 2 first in a Monte Carlo calculation,
then a high quality data collapse for $\rho_{s2}N$ can be achieved for 
$(\mu/V)_c \sim 2.3184$ and $\nu \sim 0.7$ using the data obtained with 
$\beta = 8N^{1/2}$. Our results imply that one might need to reconsider the 
established phase diagrams of some models if the accurate location of the 
critical point is crucial in obtaining a conclusion.

\end{abstract}

\maketitle


\section{Introduction}

Searching and investigating models in which one might observe a supersolid phase (SS) has been one of the central research interests 
in condensed matter physics recently. Indeed, it is reported numerically that several spin models and extended Bose-Hubbard models 
provide convincing evidence for the existence of a supersolid state, where the long range superfluid order and solid order 
coexist \cite{Sen05,Wes05,Mel05,Bon05,Mel06,Ng06,Sen07,Gan07_1,Wes07,Gan07_2,Laf07,Che08}. For example, a Monte Carlo investigation 
of bilayer spin-1/2 Heisenberg model with an external uniform magnetic field on the square lattice demonstrates that a 
field-induced supersolid phase can be stabilized when the magnitude of the external magnetic field takes certain 
values \cite{Ng06,Laf07}. A supersolid phase is observed numerically by considering interacting bosons as 
well \cite{Sen05,Wes05,Mel05,Bon05,Mel06,Gan07_1,Wes07,Gan07_2,Che08}. Further, it is expected that SS and the quantum phase 
transitions out of SS might be realized experimentally using ultracold atoms on optical lattices \cite{Gre02,Sto04,Blo06}. 
Although rich phase diagrams have been obtained for both spin models and extended Bose-Hubbard models, a detailed investigation 
of the nature of the phase transitions out of SS is available only for bilayer spin-1/2 Heisenberg model on the square lattice. 
Further, despite the fact that the bilayer spin-1/2 Heisenberg model can be mapped into the hard-core Bose-Hubbard model, it is known 
that both the hard-core Bose-Hubbard model on the square and the honeycomb lattices do not exhibit a stable SS phase in 
the $t/V$-$\mu/V$ phase diagram at a fixed $U/t$ \cite{Wes07,Bat00}. 
Here $t$ is the nearest-neighbor hopping parameter, $\mu$ is the chemical potential, $V$ is a nearest-neighbor 
repulsion, and $U$ stands for an onsite repulsion. As a consequence, it will be interesting to carry out an 
investigation on the nature of quantum 
phase transitions out of SS for the extended Bose-Hubbard model. Finally, considering the many noticeable properties of 
graphene \cite{Nov04,Nov05,Zha05,Cas09}, for which the underlying  lattice is a honeycomb lattice, as well as the fact that quantum 
fluctuations are expected to be more relevant on the honeycomb lattice due to its coordination number, in this paper we study the phase 
transition from a SS phase to a solid state of $\rho= 1/2$ for the extended Bose-Hubbard model on the honeycomb lattice using first 
principles Monte Carlo simulations. Here $\rho$ refers to the average number of bosons per lattice site. 

The nature of quantum phase transitions out of a SS to a superfluid and a solid phases have been studied quantitatively using first 
principles unbiased quantum Monte Carlo method for the bilayer spin-1/2 Heisenberg model \cite{Laf07}. Indeed it is demonstrated 
convincingly that the critical theories determined from the Monte Carlo data are consistent with the theoretical 
prediction \cite{Laf07}. For example, by applying the technique of data collapse to the observable superfluid density $\rho_s$, one 
reaches a perfect agreement between the numerical data and the theoretical 
prediction of superfluid-insulator universality class for the phase transition  
from a SS to a solid state \cite{Fis89,Ale04}. To obtain the ground-state phase diagrams of the desired models using finite 
temperature Monte Carlo algorithms, in particular to quantitatively study the nature of the phase transitions in the phase diagrams, 
one useful strategy is to scale the inverse temperature $\beta$ by the relation $\beta = cL^{z}$ in the simulations. Here in addition to
the dynamical critical exponent $z$ which is already introduced earlier, $L$ and $c$ appearing in $\beta = cL^z$ 
are the box sizes used in the simulations and a constant, respectively. Surprisingly, despite the fact that the correct 
way to determine the critical theory for a second order phase transition is to employ the relation $\beta= cL^z$ in the simulations, 
several studies seem to simply use $\beta = cL$ and ignore the impact of the dynamical critical exponent $z$ when obtaining the 
ground-state phase diagrams. Whether the strategy of using $\beta = cL$ instead of $\beta = cL^z$ in the Monte Carlo 
calculations has noticeable influence on determining the critical theory for a second order phase transition remains to explore.

The motivation of our investigation on the quantum phase transition from a supersolid state
to a solid state of $\rho = 1/2$ by varying $\mu/V$ at fixed $U/t = 20$ and $t/V=0.16$ in the parameter space
for the extended Bose-Hubbard model is twofold. First of all, we 
would like to determine the critical chemical potential $(\mu/V)_c$ as precise 
as possible since such a study is useful in calculating critical exponents
such as $\beta$ ($\beta/\nu$) and $\eta$. Secondly, since one should scale the
inverse temperature $\beta$ with the system size for the phase transition 
considered here due to the theoretical prediction $z=2$,
we would like to understand the impact
of scaling $\beta$ linearly with the linear length of the system on studying
the corresponding critical theory of this transition. Indeed as we will
demonstrate later, scaling $\beta$ linearly with the system linear length
either leads to poor data collapse for the observables measured in this 
study, or one would arrive at a different critical theory than the expected
one.

This paper is organized as follows. After a brief description of the motivation behind
this study, the extended Bose-Hubbard model considered here as well as the observables measured in 
our Monte Carlo simulations are introduced. Follows that we present our numerical results.
In particular, the corresponding critical point is determined with high precision 
by the method of data collapse. The subtlety of determining the critical theory for the transition from a SS phase to a solid state
of $\rho = 1/2$ is demonstrated as well. Finally, a section is devoted to conclude our investigation.

\section{The Model and Observables}
The extended Bose-Hubbard model considered in this study is given by
\begin{eqnarray}
\label{hamiltonian}
H&=&-t \sum_{\langle i,j \rangle} \left( b^\dagger_i b_j + b^\dagger_j b_i \right)
  +\frac{U}{2} \sum_i  n_i (n_i-1)\nonumber\\
  &&+V \sum_{\langle i,j \rangle} n_i n_j
  -\mu \sum_i n_i
,\end{eqnarray}
where $b^{\dagger}_i$ ($b_i$) are the bosonic 
creation (annihilation) operators at site $i$ and $n_i$ is the occupation number at lattice site $i$.
Further, $U$, $V$, $t$, and $\mu$ appearing in Eq.~(\ref{hamiltonian}) are defined as before.

The honeycomb lattice with periodic spatial boundary conditions implemented in 
our simulations is depicted in figure 1. 
The dashed rectangle in figure 1, which contains $4$ spins, is the elementary
cell for building a periodic honeycomb lattice covering a rectangular area.  
For instance, the honeycomb lattice shown in figure 1 contains 3 $\times$ 3 
elementary cells. The lattice
spacing $a$ is 
the distance between two neighboring sites. The honeycomb lattice is not a
Bravais lattice. 
Instead it consists of two triangular Bravais sub-lattices $A$ and $B$
(depicted by solid and open circles in figure 1). As a consequence, 
the momentum space of the honeycomb lattice is a doubly-covered Brillouin 
zone of the two triangular sub-lattices.

Following earlier works in \cite{Wes07,Gan07_2}, our primary interest for the model 
described by Eq.~(\ref{hamiltonian}) is to
study the phase transitions from a supersolid phase to a solid phase of $\rho = 1/2$ 
and from a superfluid phase
to a Mott insulator phase of $\rho = 1$ by varying the chemical potential $\mu/V$ at fixed $U/t = 20$ and 
$t/V = 0.16$ in the parameter space. 
In particular we would like to determine the critical chemical potentials $(\mu/V)_c$ as well as
the correlation length critical exponents $\nu$ and dynamical critical exponents $z$ 
for these two phase transitions accurately. 
To fulfill these tasks, we have measured the superfluid density 
$\rho_{si} = \langle W_i^2 \rangle /\beta$ with $i \in \{1,2\}$ in our simulations
and have focused on the finite-size scaling of $\rho_{s2}N^{z/2}$\symbolfootnote[1]{Strictly speaking, the observable
$\rho_{si}$ is defined
by $\rho_{si} = \langle W_i^2 \rangle /( \beta t)$. However since the parameter $t$ is fixed to be $t/V = 0.16$ in
our investigation, we will use $\rho_{si} = \langle W_i^2 \rangle /\beta $ instead since the conclusions are not 
affected.}. Here $\langle W_i^2 \rangle$ is the winding number fluctuation in $i$-direction.

\vskip0.75cm

\begin{figure}[H]
\begin{center}
  \includegraphics[width=0.5\textwidth]{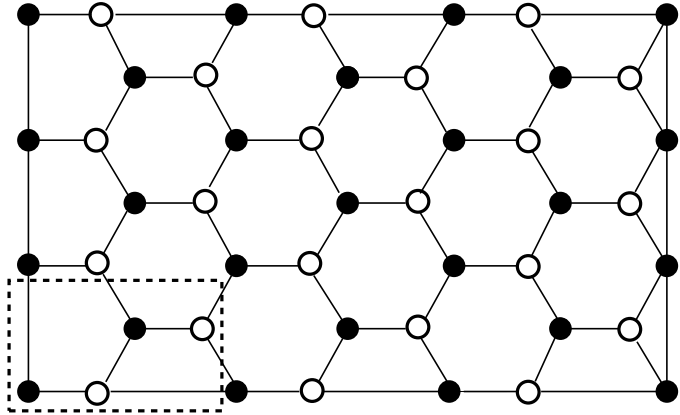}\vskip0.25cm
\end{center}\vskip-0.1cm
\caption{
The periodic honeycomb lattice consisting of two triangular
sub-lattices A and B, which are depicted by solid and open 
circles, respectively. The dashed rectangle is an elementary cell
for building a periodic honeycomb lattice covering a rectangular 
area.}
\label{fig1}
\end{figure}

\vskip0.5cm

\section{Determination of the critical point}

To determine the
location of the critical point in the parameter space $\mu/V$, one useful technique
is to study the finite-size scaling of certain observables. For example,
if the transition is second order, then near the transition, the observable 
$\rho_{si} N^{z/2}$ for $i\in \{1,2\}$ should be described well by the following finite-size scaling 
ansatz
\begin{equation}
\label{FSS}
{\cal O}_{N}(j) = g_{{\cal O}}(jN^{1/2\nu},N^{z/2}/\beta), 
\end{equation}
where ${\cal O}_{N}$ stands for $\rho_{si}N^{z/2}$, $j = (\mu_c-\mu)/\mu_c$, and
$\nu$ is the critical exponent corresponding to the correlation length $\xi$. 
Finally $g_{{\cal O}}$ appearing above is a 
smooth function of the variables $jN^{1/2\nu}$ and $N^{z/2}/\beta$. 
In writing Eq.~(\ref{FSS}), we have ignored explicitly the confluent correction to the scaling. Taking this correction into account, one concludes that for large enough $\beta$ when the finite-temperature
effects can be ignored,
the curves of different $N^{1/2}$ for ${\cal O}_{N}$, as functions of $\mu/V$,
should have the tendency to intersect at critical point $(\mu/V)_c$ for large $N^{1/2}$. 
Theoretically, it is predicted that the transition from a supersolid phase to a solid phase of $\rho = 1/2$
for this model is governed by the superfluid-insulator universality class, namely one has $\nu=0.5$, $\beta = 0.5$ 
and $z=2$ \cite{Fis89,Ale04}.
In the following we will apply the finite-size scaling formula,
Eq.~(\ref{FSS}), to the observable $\rho_{s2} L$ to 
determine $(\mu/V)_c$. In particular, we would like to examine whether the theoretical prediction is consistent with
our Monte Carlo data.

\vskip0.75cm

\begin{figure}[H]
\begin{center}
  \includegraphics[width=0.5\textwidth]{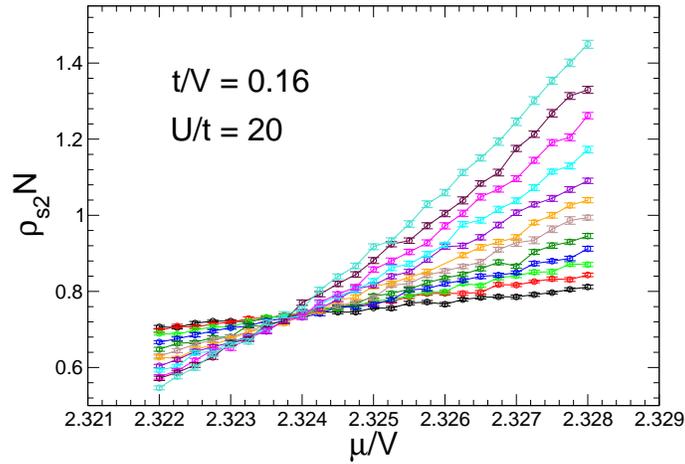}\vskip0.25cm
\end{center}\vskip-0.1cm
\caption{Monte Carlo data of $\rho_{s2}N$ as functions of $\mu/V$ 
for the quantum phase transition from a supersolid state to a solid state of 
$\rho = 1/2$. For a given fixed $N$, the
inverse temperature $\beta$ used for this set of simulations is given by $\beta = 2N^{1/2}$. }
\label{fig2}
\end{figure}

\vskip0.5cm

\begin{figure}[H]
\begin{center}
  \includegraphics[width=0.5\textwidth]{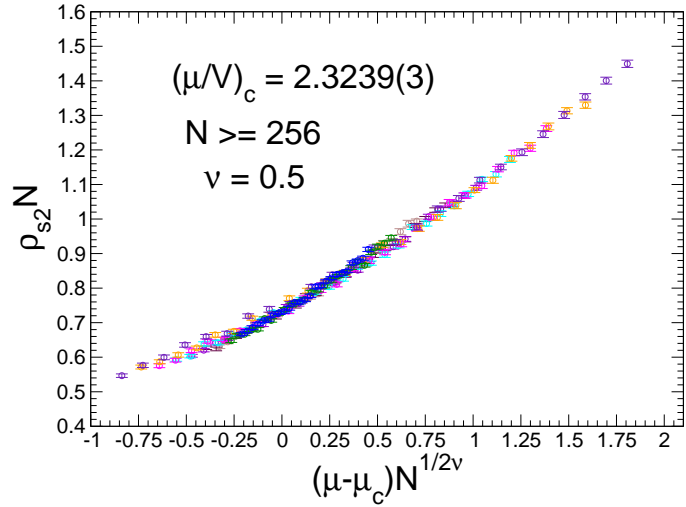}\vskip0.25cm
\end{center}\vskip-0.1cm
\caption{Data collapse for the observable $\rho_{s2}N$ with $16^2 \le N \le 32^2$
for the quantum phase transition from a supersolid state to a solid state of 
$\rho = 1/2$.
The data is calculated using $\beta = 2N^{1/2}$. In obtaining the figure, 
the critical exponent $\nu$ is fixed to be the expected theoretical value $\nu = 0.5$. }
\label{fig3}
\end{figure}

\vskip0.5cm

To study the critical theory for the phase transition of our central interest, 
we have performed
large scale Monte Carlo simulations using the directed loop algorithms available in the
ALPS library \cite{ALP07}. Without losing the generality, in our simulations we have 
fixed $V$ to be $1.0$ and have varied $\mu$. 
To determine $(\mu/V)_c$ using the finite-size scaling ansatz Eq.~(\ref{FSS}), one
needs to use large enough $\beta$ so that the finite-temperature effects
are negligible. For this purpose, we have carried out a trial simulation
at $\mu/V = 2.53$ with $N = 14^2$ and we have found that one already obtains 
the zero-temperature result for $\rho_{s2}$ using $\beta = 2 \times 14^{1/2}$.
Hence we use $\beta = 2N^{1/2}$ in other simulations as well.
Notice the strategy of applying $\beta = 2N^{1/2}$ or similar ones for a fixed $N^{1/2}$ in the simulations
was used in many studies exploring the
phase diagrams of certain models in the literature.
After determining the relation $\beta = 2N^{1/2}$ which allows one to access 
zero-temperature values for the observable $\rho_{s2}$, we have further carried out large scale simulations with
$N$ ranging from $N = 10^2$ to $N = 32^2$. Figure \ref{fig2} demonstrates the 
results of $\rho_{s2}N$ as functions of $\mu/V$ for this set of simulations. 
The figure indicates that the phase transition is likely a second order transition
because the curves of different $N^{1/2}$ have the tendency to intersect at a
particular $\mu/V$ is the parameter space. Using fixed $z=2$ and $\nu = 0.5$,
the best result of data collapse for $\rho_{s2}N$ is reached with 
$(\mu/V)_c = 2.3239(3)$ for the data of lattice sizes $16^2 \le N \le 32^2$ (figure \ref{fig3}). 
The quality of data collapse shown in figure \ref{fig3} is not good, but 
acceptable. One might attribute the poor quality shown in figure \ref{fig3}
to a correction in Eq.~(\ref{FSS}) that is not taken into account in our 
analysis. At this stage, one would naturally conclude that our Monte Carlo
data is consistent with the theoretical prediction, namely the critical
exponent $\nu$ and dynamical critical exponent $z$ of the phase transition
considered above is governed by $\nu=0.5$ and $z=2$.

\vskip0.75cm

\begin{figure}[H]
\begin{center}
  \includegraphics[width=0.5\textwidth]{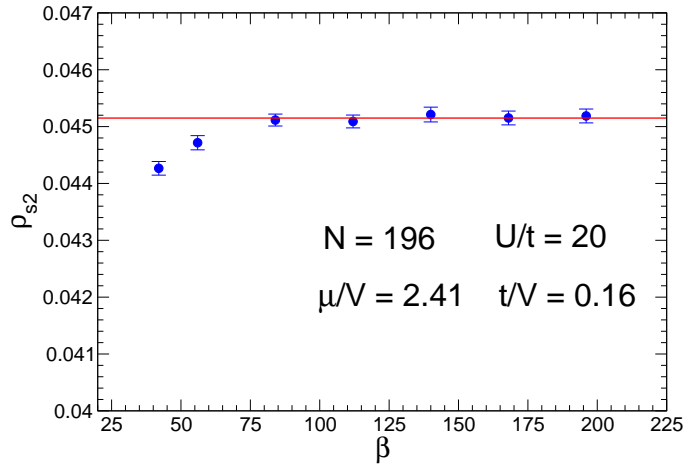}\vskip0.25cm
\end{center}\vskip-0.1cm
\caption{$\rho_{s2}$ as a function of $\beta$ at $\mu/V = 2.41$ and $N = 196$.}
\label{fig4}
\end{figure}

\vskip0.5cm

\begin{figure}[H]
\begin{center}
  \includegraphics[width=0.5\textwidth]{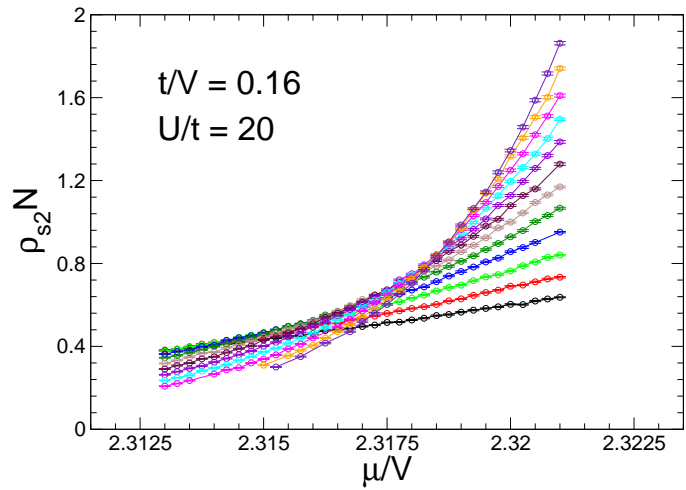}\vskip0.25cm
\end{center}\vskip-0.1cm
\caption{Monte Carlo data of $\rho_{s2}N$ as functions of $\mu/V$
for the quantum phase transition from a supersolid state to a solid state of 
$\rho = 1/2$. For a given fixed $N$, the
inverse temperature $\beta$ used for this set of simulations is given by $\beta = 8N^{1/2}$.}
\label{fig5}
\end{figure}

\begin{figure}[H]
\begin{center}
  \includegraphics[width=0.5\textwidth]{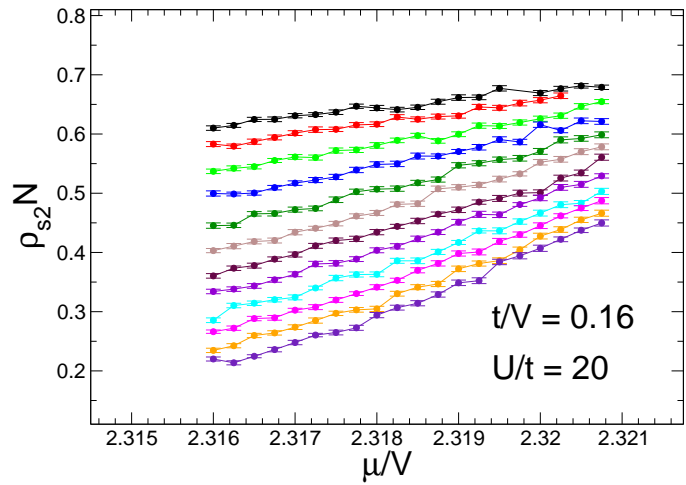}\vskip0.25cm
\end{center}\vskip-0.1cm
\caption{Monte Carlo data of $\rho_{s2}N$ as functions of $\mu/V$ 
in the range $2.316 \le \mu \le 2.32075$ with $10^2 \le N \le 32^2$. 
For a given fixed $N$, the
inverse temperature $\beta$ used for this set of simulations is given by $\beta = 2N^{1/2}$.}
\label{fig6}
\end{figure}

To make sure that we indeed obtain the ground-state properties of the model, 
we repeat above analysis by firstly determining the required $\beta$ 
for reaching the ground-state value of $\rho_{s2}$ at $\mu/V = 2.41$ and
$N = 14^2$.  
Surprisingly, we find that one has to use $\beta \sim 6N^{1/2}$ in order to
reach the zero-temperature value of $\rho_{s2}$ (figure \ref{fig4}). Because
of this observation,
we have performed another set of simulations using $\beta = 8N^{1/2}$.
The results of $\rho_{s2}N$ as functions of $\mu/V$ for this new set of runs
is shown in figure \ref{fig5}. By comparing figures \ref{fig2} and 
\ref{fig5}, one clearly observes a statistically difference between the
critical chemical potentials calculated from these two set of data. To make the 
discrepancy between these two critical chemical potentials
more transparent, we additionally simulating the model using 
$\beta = 2N^{1/2}$ in the range $2.316 \le \mu/V \le 2.32075$ where
$(\mu/V)_c$ for the second set of data (which are determined with $\beta = 8N^{1/2}$) 
is located. Figure
\ref{fig6} shows the results of $\rho_{s2}N$ as functions of 
$\mu/V$ for these new runs. No intersection between the curves of different 
$N^{1/2}$ shown in figure \ref{fig6} confirms our observation that the 
$(\mu/V)_c$ for these two set of data are statistically different.

\vskip0.75cm

\vskip0.5cm

\vskip0.5cm

\begin{figure}[H]
\begin{center}
  \includegraphics[width=0.5\textwidth]{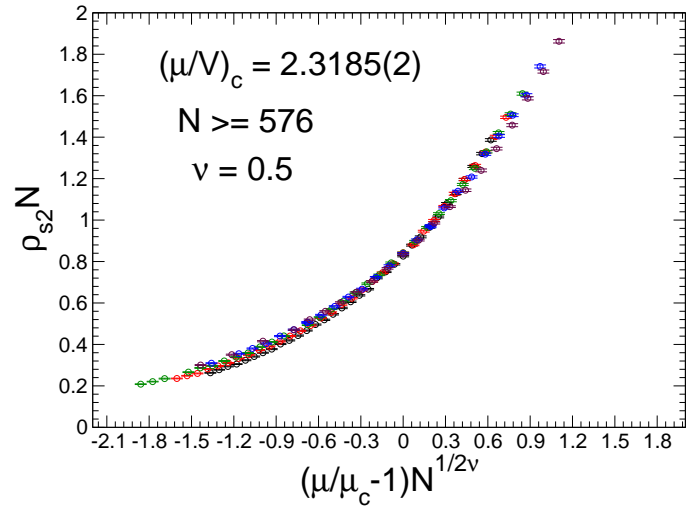}\vskip0.25cm
\end{center}\vskip-0.1cm
\caption{Data collapse of the observable $\rho_{s2}N$ with $24^2 \le N \le 32^2$
for the quantum phase transition from a supersolid state to a solid state of 
$\rho = 1/2$. The data is 
calculated using $\beta = 8N^{1/2}$. In obtaining the figure, the critical exponent $\nu$ 
is fixed to be the expected theoretical value $\nu = 0.5$.}
\label{fig7}
\end{figure}

\begin{figure}[H]
\begin{center}
  \includegraphics[width=0.5\textwidth]{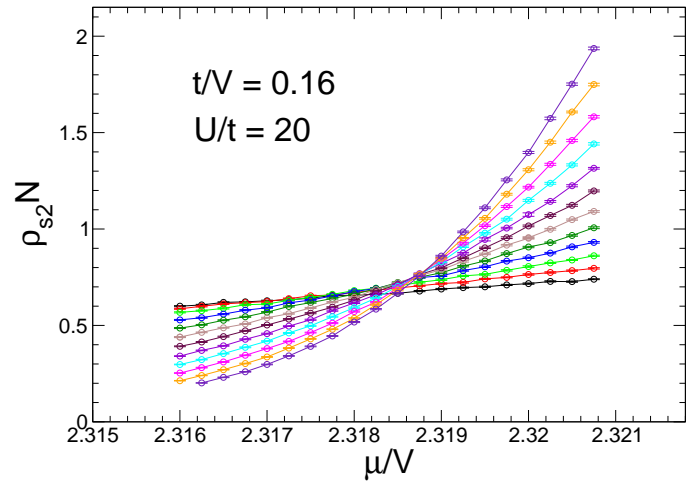}\vskip0.25cm
\end{center}\vskip-0.1cm
\caption{Monte Carlo data of $\rho_{s2}N$ as functions of $\mu/V$
for the quantum phase transition from a supersolid state to a solid state of 
$\rho = 1/2$. For a given fixed $N$, the
inverse temperature $\beta$ used for this set of simulations is given by $\beta = N/2$.}
\label{fig8}
\end{figure}

\begin{figure}[H]
\begin{center}
  \includegraphics[width=0.5\textwidth]{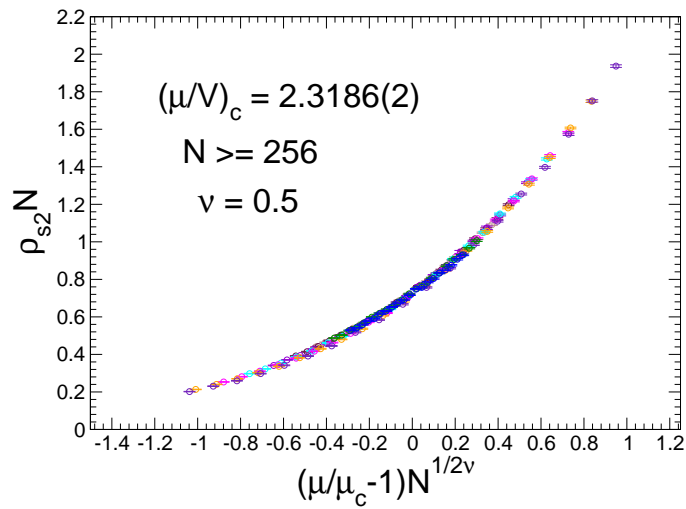}\vskip0.25cm
\end{center}\vskip-0.1cm
\caption{Data collapse of the observable $\rho_{s2}N$ with $16^2 \le N \le 32^2$
for the quantum phase transition from a supersolid state to a solid state of 
$\rho = 1/2$. The data is 
calculated using $\beta = N/2$. In obtaining the figure, the critical exponent $\nu$ 
is fixed to be the expected theoretical value $\nu = 0.5$.}
\label{fig9}
\end{figure}

After demonstrating that the critical chemical potentials determined from the sets of 
data obtained using $\beta = 2N^{1/2}$ and $\beta = 8N^{1/2}$ are statistically
different, let us return to the analysis of the second set of $\rho_{s2}$ data. 
Yet another surprise we find is that  
a good data collapse for $\rho_{s2}N$ can hardly be achieved 
if data points 
of small $N$ are included in the analysis. Only with data of large $N$, 
namely $N \ge 24^2$, a reasonable data collapse for $\rho_{s2}N$ with a
fixed $\nu=0.5$ can be obtained given that $(\mu/V)_c = 2.3185(2)$ 
(figure \ref{fig7}). The results presented so far in this study raise an
interesting question, namely what values of $\beta$ should be used in our simulations
in order to obtain the ground-state values for the observable $\rho_{s2}$.
In other words, should one perform another set of simulations with even 
larger values of $\beta$? Actually by examining the relevant
finite-size scaling ansatz Eq.~(\ref{FSS}) carefully, one would realize 
that the correct strategy is to scale $\beta$ with $N$. Figure
\ref{fig8} shows the results of $\rho_{s2}N$ obtained 
with $\beta = N/2$ as functions of $\mu/V$. The tendency 
of crossing between curves of different $N^{1/2}$ shown in figure \ref{fig8}
is much stronger compared to those presented in figures \ref{fig2} 
and \ref{fig5}. Finally the quality of data collapse for $\rho_{s2}N$ demonstrated in figure \ref{fig9}
which are obtained from the set of simulations 
using $\beta = N/2$ is also much better than those found in figures \ref{fig3} and \ref{fig7}. 
Notice the $(\mu/V)_c$ determined from the data simulated with $\beta = N/2$ is given by $(\mu/V)_c = 2.3186(2)$
which agrees with $(\mu/V)_c = 2.3185(2)$, but is statistically different from $(\mu/V)_c = 2.3239(3)$ obtained earlier 
from the data determined using $\beta = 2N^{1/2}$.   

\vskip0.75cm

\vskip0.5cm

\vskip0.5cm

\vskip0.5cm

\begin{figure}[H]
\begin{center}
  \includegraphics[width=0.5\textwidth]{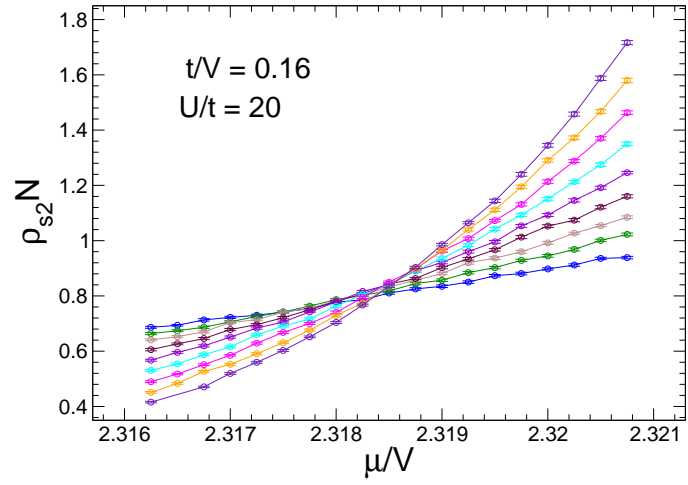}\vskip0.25cm
\end{center}\vskip-0.1cm
\caption{Monte Carlo data of $\rho_{s2}N$ as functions of $\mu/V$ for the 
quantum phase transition from a supersolid state to a solid state of 
$\rho = 1/2$. For a given fixed $N$, the
inverse temperature $\beta$ used for this set of simulations is given by $\beta = N/4$.}
\label{fig10}
\end{figure}

\begin{figure}[H]
\begin{center}
  \includegraphics[width=0.5\textwidth]{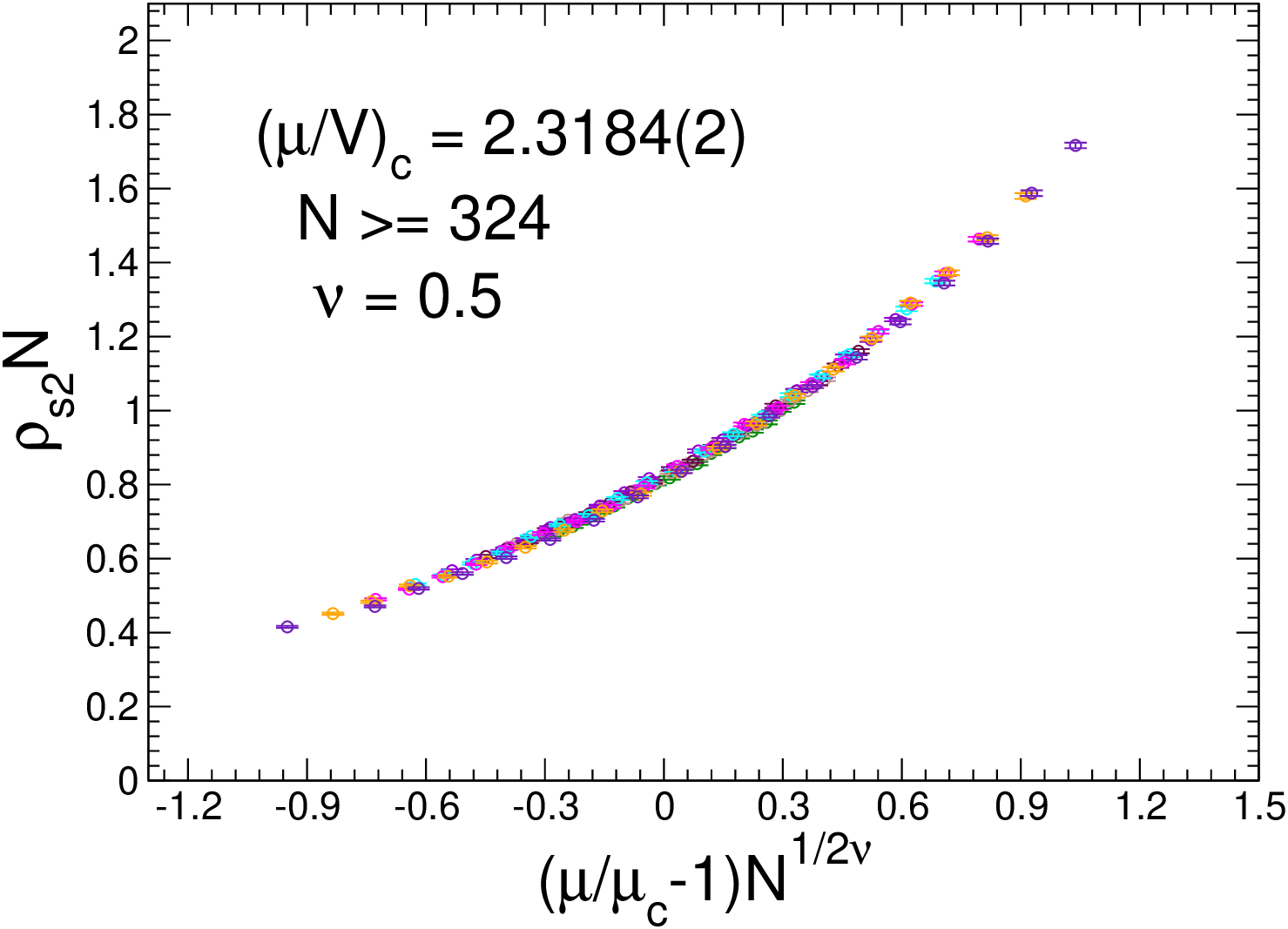}\vskip0.25cm
\end{center}\vskip-0.1cm
\caption{Data collapse of the observable $\rho_{s2}N$ with $18^2 \le N \le 32^2$
for the quantum phase transition from a supersolid state to a solid state of 
$\rho = 1/2$. The data is 
calculated using $\beta = N/4$. In obtaining the figure, the critical exponent $\nu$ 
is fixed to be the expected theoretical value $\nu = 0.5$.}
\label{fig11}
\end{figure}

\begin{figure}[H]
\begin{center}
  \includegraphics[width=0.5\textwidth]{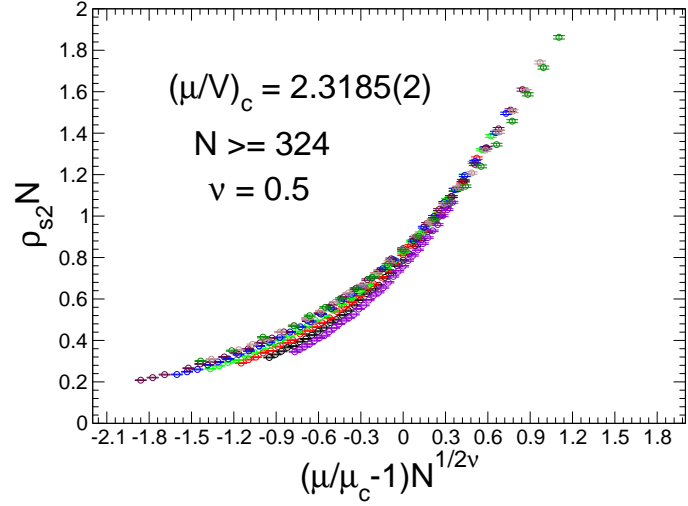}\vskip0.25cm
\end{center}\vskip-0.1cm
\caption{Data collapse of the observable $\rho_{s2}N$ with $18^2 \le N \le 32^2$ for the 
quantum phase transition from a supersolid state to a solid state of 
$\rho = 1/2$. The data is 
calculated using $\beta = 8N^{1/2}$. In obtaining the figure, the critical exponent $\nu$ 
is fixed to be the expected theoretical value $\nu = 0.5$.}
\label{fig12}
\end{figure}

\begin{figure}[H]
\begin{center}
  \includegraphics[width=0.5\textwidth]{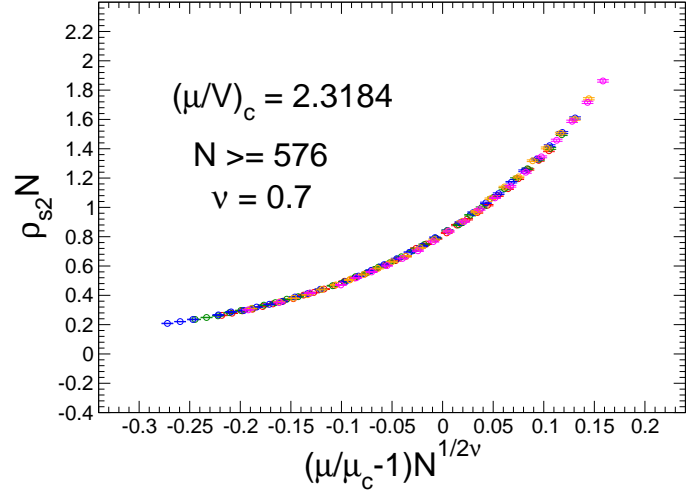}\vskip0.25cm
\end{center}\vskip-0.1cm
\caption{Data collapse of the observable $\rho_{s2}N$ with $24^2 \le N \le 32^2$ for the 
quantum phase transition from a supersolid state to a solid state of 
$\rho = 1/2$. The data is 
calculated using $\beta = 8N^{1/2}$. In obtaining the figure, the critical chemical potential
$\mu_c$ and the critical exponent $\nu$ 
are fixed to be $\mu_c = 2.3184$ and $\nu = 0.7$, respectively.}
\label{fig13}
\end{figure}

Of course, one might argue that since $N/2 > 8N^{1/2}$ in the range of 
$N$ used in the simulations and analysis, the better results shown in
figures \ref{fig8} and \ref{fig9} than those in figures 
\ref{fig2}, \ref{fig3}, \ref{fig5}, and 
\ref{fig7} is simply because one reaches the ground-state values of 
$\rho_{s2}$ for the set of data determined using $\beta = N/2$ and the 
$\rho_{s2}$ data points calculated with $\beta = 8N^{1/2}$ still receive 
finite-temperature effects. To rule out such a possibility, we have performed another
set of simulations using $\beta = N/4$ with which the condition $N/4 \le 8N^{1/2}$ 
is satisfied for 
$N^{1/2} = 18, 20,..., 32$. Figures \ref{fig10} and \ref{fig11} show the results
of $\rho_{s2}N$ with $18^2\le N \le 32^2$ for these new simulations using $\beta = N/4$ as functions
of $\mu/V$ and the corresponding results of data collapse. 
In obtaining figure \ref{fig11}, the critical exponent $\nu$ is fixed to $\nu=0.5$ as before. 
Again the quality of crossing and
data collapse seen in figures \ref{fig10} and \ref{fig11} are much better
than those in figures \ref{fig5} and \ref{fig7}. To make the comparison on the same footing,
figure \ref{fig12} shows the results of data collapse for $\rho_{s2}N$ 
determined using $\beta = 8N^{1/2}$ with $18^2 \le N \le 32^2$. 
Figures \ref{fig11} and \ref{fig12} clearly indicate that
the quality of data collapse in figure \ref{fig11} is much better than that of figure \ref{fig12}.
Interestingly, assuming
$z = 2$ is determined first in a Monte Carlo calculation before one performs the large scale simulations for obtaining
$\rho_{s2}$ using $\beta = 8N^{1/2}$, then a good data collapse 
can be reached with $(\mu/V)_c = 2.3184$ and $\nu = 0.7$ using the data of $\rho_{s2}N (N \ge 24^2)$ calculated
with $\beta = 8N^{1/2}$ (figure \ref{fig13}). While $(\mu/V)_c = 2.3184$ is consistent with 
$(\mu/V)_c = 2.3185(2)$ and $(\mu/V)_c = 2.3186(2)$, the $\nu = 0.7$ we find is significantly different 
from the theoretical expectation $\nu=0.5$. Without the theoretical input 
$\nu=0.5$, one might be misled by figure \ref{fig13} to conclude an 
unconventional phase transition is observed for this model. Our results 
presented in this study clearly imply the importance of using the correct 
relation for $\beta$ and $N^{1/2}$ in investigating the critical theory of a second 
order phase transition.  

\vskip0.75cm

\vskip0.5cm

\vskip0.5cm

\vskip0.5cm

\vskip0.5cm

\begin{figure}[H]
\begin{center}
  \includegraphics[width=0.5\textwidth]{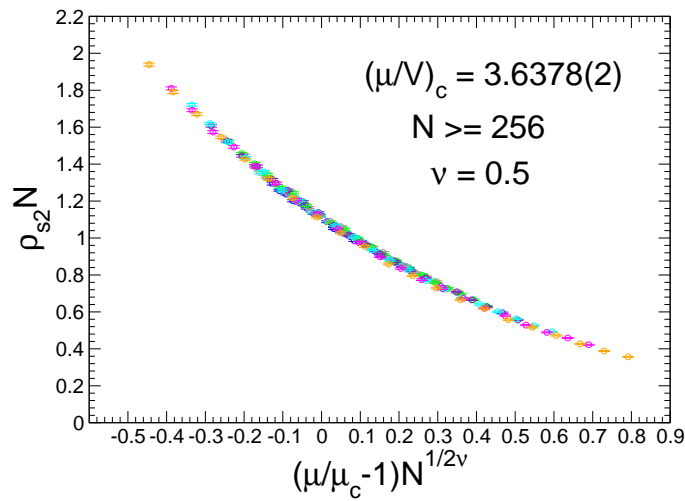}\vskip0.25cm
\end{center}\vskip-0.1cm
\caption{Data collapse of the observable $\rho_{s2}N$ with $18^2 \le N \le 32^2$ for the 
quantum phase transition from a superfluid state to a Mott insulator state of 
$\rho = 1$. The data is 
calculated using $\beta = N/2.5$. In obtaining the figure, the critical exponent $\nu$ 
is fixed to be the expected theoretical value $\nu = 0.5$.}
\label{fig14}
\end{figure}

In additional to the quantum phase transition from a supersolid phase to
a solid phase of $\rho = 1/2$ for the extended Bose-Hubbard model, we have
studied the critical theory of the phase transition from a superfluid phase
to a Mott insulator phase of $\rho = 1$ as well. Inspired by the subtlety we 
observed
earlier, we use $\beta = N/2.5$ for this new investigation since the dynamical
critical exponent $z$ for this transition is predicted to be 2 theoretically.
Indeed using $\nu=0.5$ which is the expected theoretical value for $\nu$,
a good data collapse is reached for $\rho_{s2}N$ provided that $(\mu/V)_c = 3.6378(2)$
and $N \ge 16^2$ (figure \ref{fig14}). In other words, our Monte Carlo
data is fully compatible with the predicted universality class for the transition 
from a superfluid phase to a Mott insulator phase of $\rho = 1$.

\vskip0.75cm

\section{Discussions and Conclusions}
In this paper, we use first principles Monte Carlo methods to study the quantum
phase transitions from a supersolid phase to a solid phase of $\rho=1/2$ and 
from a superfluid phase to a Mott insulator phase of $\rho = 1$ by varying the chemical potential $\mu/V$ 
at fixed $U/t=20$ and $t/V=0.16$ in the parameter space for the extended 
Bose-Hubbard model. We confirm that our Monte 
Carlo results for both phase transitions are fully compatible with the 
corresponding theoretical prediction. Specifically, we obtain a good quality 
of data collapse
for $\rho_{s2}N$ with fixed $\nu=0.5$ and $z=2$ in the analysis. On the other
hand, we find it is subtle to determine the correct critical chemical
potential $(\mu/V)_c$ for the phase transition from a supersolid phase to
a soild phase of $\rho=1/2$. For example, while the $(\mu/V)_c$ calculated
from a set of $\rho_{s2}$ data obtained using $\beta = 2N^{1/2}$ is given
by $(\mu/V)_c = 2.3289(2)$, another
set of $\rho_{s2}$ data determined from the simulations with $\beta = N/2$ leads 
to $(\mu/V)_c = 2.3186(2)$. The deviation of these two values of 
$(\mu/V)_c$ found from different strategies of scaling the inverse 
temperature $\beta$ with the system sizes is statistically significant.
The later result for $(\mu/V)_c$ should be
more reliable considering the fact that the dynamical critical exponent 
$z$ is 2. We also demonstrate that using the incorrect scaling of $\beta$, 
namely one scales $\beta$ linearly with the system linear length $N^{1/2}$, 
would either lead to bad quality of data collapse for the observables 
considered here, or one would arrive at a different critical theory for 
the phase transition investigated in this study. Despite the fact that
the principles behind the technique of data collapse is exact, it might 
mislead one to conclude an observation of an unconventional 
phase transition. Our results of $(\mu/V)_c = 2.3184$ and $\nu=0.7$ obtained
by carrying out the data collapse for $\rho_{s2}N$ determined from the 
simulations with $\beta = 8N^{1/2}$ is a good example of such a scenario. 
When applying the technique of data collapse to study the critical theory of a 
second order phase transition, it might 
be useful to employ this technique in a more rigorous manner. 
At this stage, it is a surprise that the motivation behind our investigation, 
namely whether scaling $\beta$ linearly with the system linear length for
studying a second order phase transition with $z=2$ is an appropriate strategy, is
ignored in some studies of phase diagrams for certain models in the literature. 
To re-examine the
corresponding critical points seems to be the required step in order to confirm the quantitative correctness of
the established phase diagrams of some spin and bose models.
Our results does not necessarily indicate that the phase diagram of the
bilayer spin-1/2 Heisenberg model in an external magnetic field studied in \cite{Laf07} 
and the phase diagram of extended Bose-Hubbard model mapped out in \cite{Gan07_2} 
are not quantitatively correct. However one might need to reconsider the phase
diagrams of these models if the precise location of a critical point in the parameter space
is crucial in drawing a conclusion. Finally, based on the results presented in this paper, in particular the
unexpected $\nu = 0.7$ obtained from the data 
$\rho_{s2}N$ calculated with $\beta = 8N^{1/2}$, it will be important to determine
the dynamical critical exponent $z$ first when investigating the critical theory for a second order phase transition which might
be described by a new universality class and is poor known theoretically.

\section*{ACKNOWLEDGMENTS}
Partial support from NCTS (North) and NSC (F.J.J.) as well as SNF (U.G.) is acknowledged.
The ``Albert Einstein Center for Fundamental Physics'' at Bern 
University is supported by the ``Innovations- und Kooperationsprojekt C-13'' of
the Schweizerische Universit\"atskonferenz (SUK/CRUS).

\end{document}